\documentstyle[aps,epsf,tighten]{revtex}
\def\beq{\begin{equation}}
\def\eeq{\end{equation}}
\def\bea{\begin{eqnarray}}
\def\eea{\end{eqnarray}}

\def\eqlab#1{\label{eq:#1}}
\def\figlab#1{\label{fig:#1}}

\def\eref#1{(\ref{eq:#1})}
\def\eqref#1{eq.~(\ref{eq:#1})}
\def\Eqref#1{Eq.~(\ref{eq:#1})}

\def\Figref#1{Fig.~\ref{fig:#1}}

\def\sla#1{#1 \hspace{-2.4mm} \slash}

\def\half{\mbox{\small{$\frac{1}{2}$}}}

\def\al{\alpha}

\def\ga{\gamma} 
 \def\De{{\it\Delta}}
  \def\eps{\epsilon}

 \def\La{{\it\Lambda}}

\def\si{\sigma} \def\Si{{\it\Sigma}}
\def\th{\theta}  
\def\w{\omega}

\def\dd{{\rm d}}

\def\ie{{i.e.\ }}

\def\CF#1#2#3#4{#1 {#2} (#3) #4}  
\def\ibid {{\it ibid.\ }}

\def\ncim {Nuovo~Cim.}
\def\np {Nucl.~Phys.}
\def\prev {Phys.~Rev.}
\def\prc {Phys.~Rev.~C}
\def\prd {Phys.~Rev.~D}
\def\plett {Phys.~Lett.}

\def\rk{{\rm k}}

\def\N{{\rm N}}
\def\preprint#1{\hfill {\rm\small #1}}

\begin{document}


\title{\preprint{THU-97/32}\\[5mm]
Relativistic covariance of quasipotential equations}
\author{V.~Pascalutsa and J.A.~Tjon}
\address{Institute for Theoretical Physics, University of Utrecht,
3584 CC Utrecht, The Netherlands}
\date{November 3, 1997}

\maketitle
\thispagestyle{empty}
\medskip

\begin{abstract}
We argue that most of the relativistic 3-D (quasipotential) equations
used in hadron physics are inconsistent with the discrete symmetries like
charge conjugation and CPT, yielding an incorrect Lorentz structure
for the calculated Green's functions. An exception to this is 
the equal time approximation to the Bethe-Salpeter equation.
We present a covariant quasipotential model for the pion-nucleon
interaction based on hadronic degrees of freedom and satisfying the
full covariance.
\vspace{2mm}

\noindent
{\it PACS number(s)}: {11.10.St, 11.30.Er, 21.45.+v, 13.75.Gx} 
\vspace{1mm}


\end{abstract}
\vspace{7mm}

In recent years there has been a considerable
interest to formulate 3-dimensional (3-D) dynamical equations
for the $\pi N$ system satisfying relativistic covariance \cite{PeJ91,GrS93}.
Within a relativistic field theory framework
the Bethe-Salpeter (BS) equation,
see \Figref{bsef}, can conveniently be used
to carry out a manifestly covariant reduction to a 
3-D quasipotential (QP) equation~\cite{BSLT,Tod73,Grs82,MaW87,tlh}.
This reduction involves an assumption about
the singularities of the BS kernel after which the integration
over the 0-th component (time or energy) can easily be done explicitly.
There is, however, no unique scheme for the choice of the 
QP equation. The covariant reduction can be
done in infinitely many different ways.
Certainly, one would like to restrict the choice not only by fitting 
to experimental data, some restrictions can come from various symmetries and
consistency requirements, such as the correct low-energy limit,
the correct one-body limit of the equation \cite{Grs82}, etc.
In this Letter we discuss the constraints due to discrete
Lorentz symmetries. Within a framework satisfying these symmetries,
we construct a relativistic dynamical model of the $\pi N$ interaction.

In considering existing relativistic 3-D equations,
we find that they in general do not have the correct Lorentz structure. For example, the calculated self-energy of a
spin-1/2 particle {\em does not} have the following form (required by the full Lorentz covariance),
\beq
\eqlab{form}
\Si (\sla{P})= \sla{P} A(P^2) + B(P^2),
\eeq
where $A$ and $B$ are scalar functions of the invariant $P^2$ only.
At first this result is rather surprising, naively we would expect
form \eref{form} as a natural outcome of any covariant formalism.

\begin{figure}[t]
\epsffile{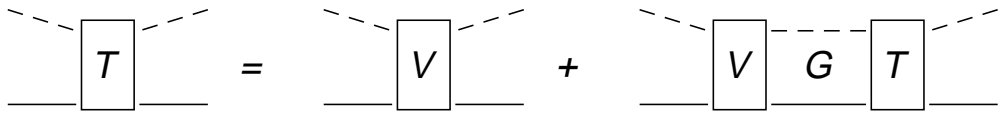}
\caption[F1]{Bethe-Salpeter equation for the $\pi N$ scattering amplitude}
\figlab{bsef}
\end{figure}

The violation of the Lorentz structure is intimately related
to the way the singularities of the BS kernel are treated
in doing the QP reduction.
To see this, let us consider a simple example of the
scalar self-energy, given by
\beq
\eqlab{sif}
\Si_S(P^2)=
i\int\! \frac{\dd^4{k}}{(2\pi)^4}\,
\frac{\Phi (k^2,P^2,P\cdot k)}{[(\half P-k)^2 - m^2-i\eps]\,
[(\half P+k)^2 - m^2-i\eps]},
\eeq
where $P$ is the relevant 4-vector, $\Phi$ is some ``interaction function''
which corresponds to the product of the two vertex functions, and which
may also have some particle propagation poles.

We immediately see that \Eqref{sif} is a function of $P^2$ only:
a sign change of $P$ can be absorbed by a change of the
loop variable $k$ in \Eqref{sif} to $-k$. In a QP 
description this
substitution in general cannot be applied in
view of the constraint in $k_0$. We can readily convince
ourself that in that case $\Si_S$ is in general not 
an even function of $P_0$ anymore.
Consider the poles of the integrand of \Eqref{sif} in
the complex $k_0$ plane. There are four poles (two in the upper
and two in the lower half-plane) coming from the propagators in
the 2-particle Green function:
$$ k_0 = \pm \half P_0 - \sqrt{m^2 + (\half \vec{P} \mp \vec{k})^2} +i\eps
\,\,\, {\rm and} \,\,\,
 k_0 = \pm \half P_0 + \sqrt{m^2 + (\half \vec{P} \mp \vec{k})^2} -i\eps. $$
We see that with a  sign change of $P_0$ and $k_0$ the
poles of the upper half-plane are interchanged with the poles
of the lower half-plane. The same symmetry exists for the
singularities of $\Phi$. Therefore, in order for $\Si_S$ to be 
even in $P_0$
the integration over $k_0$ must be independent of the choice
of the half-plane where the contour is closed. In doing a 3-D reduction,
however, one usually neglects the contribution of certain
poles, hence the result becomes dependent on the way the contour
is closed, and $\Si_S$ is not anymore an even  function of $P_0$
and consequently it cannot depend only on $P^2$.
In this sense the Lorentz structure of the self-energy is violated.

From this example we can see, that in order to simplify the
singularity structure, preserving the full Lorentz covariance,
one should first {\em remove} the `unwanted' poles, and not merely neglect them
in doing the reduction.  The so-called equal-time (ET)
approximation~\cite{tlh}
precisely provides a procedure for {\em removing} the
singularities in the $k_0$ variable. Using
the ET prescription we thus may get rid of as many poles as we like
(usually these are poles of the
vertex functions) and then perform the $k_0$ integration exactly.
For example, in the case of the scalar self-energy, the standard
ET procedure would be to remove the poles of $\Phi$ by
introducing  a quasipotential approximation to the interaction function: 
\beq
\Phi_{QP}(\tilde{k}^2, P^2) = \Phi (\tilde{k}^2,P^2,P\cdot\tilde{k})
\eeq
with $\tilde{k}=k-P(P\cdot k)/P^2.$ The self-energy becomes in this
approximation
\beq
\eqlab{sigeq}
\Si_S(P^2) \stackrel{ET}{=}
i\int\! \frac{\dd^4{k}}{(2\pi)^4}\,
\frac{\Phi_{QP}(\tilde{k}^2, P^2)}{[(\half P-k)^2 - m^2-i\eps]\,
[(\half P+k)^2 - m^2-i\eps]}
\eeq
Due to the manifestly covariant form of \Eqref{sigeq}, $\Sigma_S$
is clearly a function of $P^2$ only. 
In the c.m.\ frame ($\vec{P}=0$) \Eqref{sigeq} has a particularly
simple form. Then the interaction function 
$\Phi_{QP}(-\vec{k}^2, P^2)$ is
independent of $k_0$ and hence the
only remaining singularities of \Eqref{sigeq} in the $k_0$ plane are 
the poles of the 2-particle Green function. 
Working out the integral over $k_0$ leads to the 3-D formulation.

Obviously, similar arguments apply for the spin-1/2 particle
self-energy. Consider the dressed nucleon propagator given by
\beq
\eqlab{dprop}
S(\sla{P}) = \left[ \sla{P} - m - \Si (\sla{P}) -i\eps \right]^{-1},
\eeq
where $\Si (\sla{P})$ is the self-energy.
For simplicity we work in the  c.m.\ frame, where $P=(P_0,\,\vec{0})$.
In this frame the Dirac structure of the self-energy can simply
be represented as
\beq
\eqlab{self}
\Si(P_0)=\Si_+(P_0)\ga_+ + \Si_-(P_0) \ga_-,
\eeq
where
$\gamma_+=\half (I + \ga_0), \,\, \gamma_-=\half (I - \ga_0).$
A similar decomposition holds for the propagator,
\beq
\eqlab{propcms}
S(P_0) = S^{(+)}(P_0)\ga_+ + S^{(-)}(P_0) \ga_-
\eeq
with $S^{(\pm)}(P_0) = \pm [ P_0 \pm (- m - \Si_\pm (P_0)
+i \epsilon)]^{-1}$.
Obviously, $S^{(+)}$ corresponds to the positive and $S^{(-)}$ to the
negative energy-state propagation.

It is easy to see that, if the self-energy can be written in the
general covariant form \eref{form}, then
the following identity holds,
\beq
\eqlab{posneg}
\Si_{r} (P_0) = \Si_{-r} (-P_0),\,\,\,\, r=\pm 1
\eeq
and vise versa (in the c.m.\ frame).
 Having made this connection, we may easily
test numerically \Eqref{posneg} to see whether the form \eref{form}
 is preserved in a given dynamical model.

As was briefly reported \cite{PaT97}, we develop
a unitary dynamical model for the $\pi N$
interaction based on the solution of the BS equation in the ET approximation.
For the `driving force' we take the tree-level $\rho$-, 
$\si$-meson and $N,\, N^{*}(1470),\, \De(1232)$-baryon 
exchanges, see \Figref{potf}. 
The $\De$ and $N^{*}$ are 
widthless in the driving force, their one-pion-nucleon decay 
width is then generated dynamically in the calculation. 
In contrast to Ref.\ \cite{GrS93} no static
approximation has been made.
Note that, although the potential is crossing symmetric, the kernel of the equation and thus the resulting
amplitude is not.

\begin{figure}[p]
\epsffile{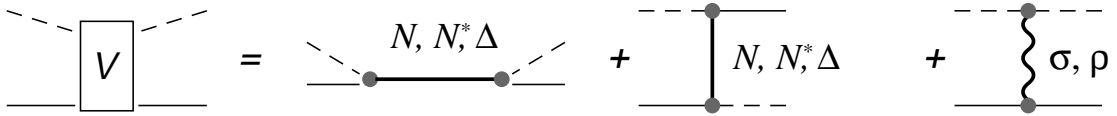}
\caption{The driving force in the $\pi N$ model.}
\figlab{potf}
\end{figure}

The bare vertices and propagators are
taken from an effective Lagrangian of the meson and the isobar fields.
We allow the pion field to be coupled only through a 
derivative coupling, which directly provides the correct low-energy limit,
at least at the tree level. The rescattering can in principle violate
the low-energy limit (in our model this may come due to
the lack of crossing symmetry). We have checked numerically that these
violations are small in our model by running a similar test as in
 Ref.\cite{PeJ91}. 
For the propagator of the $\De$ we use the Rarita-Schwinger propagator,
and the $\pi N\De$ coupling is taken to be of the general structure
determined by a coupling  constant $f_{\pi N\De}$ and an off-shell parameter
$z$, cf.\ Ref.\cite{NEK71}.
The $\rho$ exchange generates the isovector-vector current.
The strength of its coupling is rather close to that determined
by the Kawarabayashi-Suzuki relation ($g_{\pi NN}$ which
we use would imply $g_{\rho NN}g_{\rho\pi\pi}/4\pi \approx 2.8$). The $\rho$
thus mainly plays the role of the $\pi N$ contact term required
in the non-linear realizations of chiral symmetry.
The $\si$ meson
is included so as to simulate the isoscalar-scalar contribution
of the correlated two-pion exchange. Therefore, for the $\si\pi\pi$
coupling constant we take the sign determined in Ref.\cite{SDH94}, and we
find that this choice is preferable for the proper description of
the phase-shifts.

For each particle we have used a form factor depending
on the 4-momentum squared of the particle.
For a meson we take the one-boson-exchange form factor, and
for a baryon we use the form factor of Pearce and Jennings\cite{PeJ91} (with
$n_\al=2$).
The corresponding cutoff masses and other model parameters
were fitted to the $\pi N$ phase-shifts.
Their values are given in Table 1. Our fit to the S- and P-waves up
to 400 MeV pion kinetic energy is presented
in \Figref{pinf}. 

\begin{figure}[h]
\epsffile{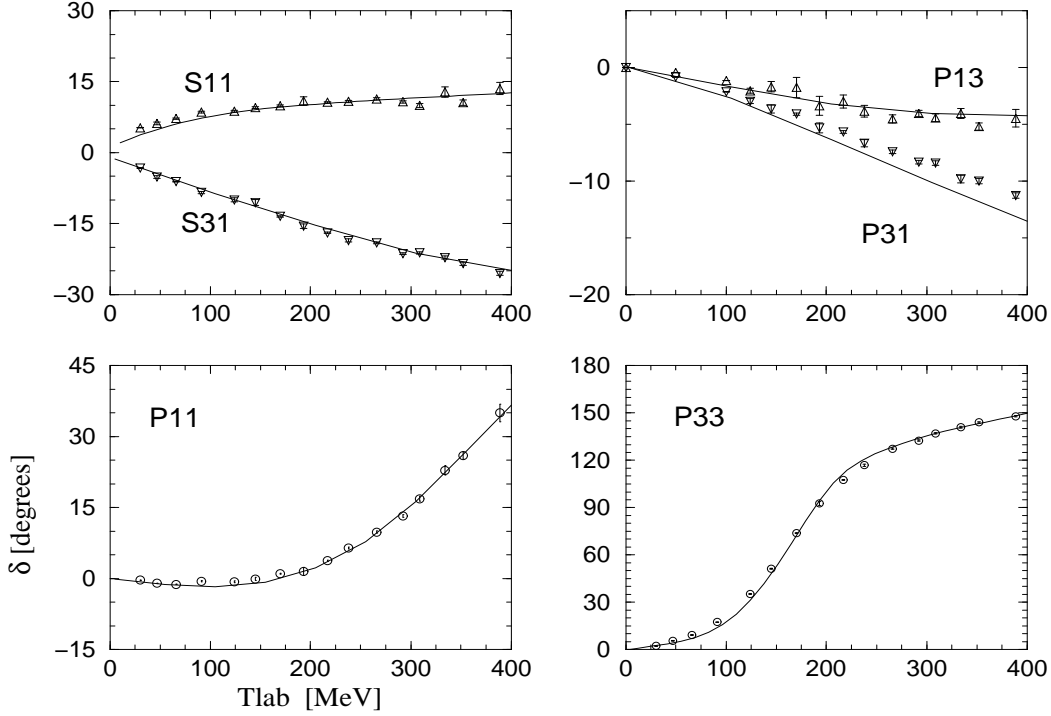}
\vskip2mm
\caption[F4]{The S- and P-wave $\pi N$ phase-shifts. 
Solid lines represent our calculation,
data points are from the SM95 partial-wave analysis\cite{ASW95}. }
\figlab{pinf}
\end{figure}
 
\begin{table}[b]
\begin{tabular}{cc}
{\rm coupling constants} & {\rm masses [GeV]}\\ \hline
$g_{\pi NN}^2/4\pi = 13.5$ & $\La_N = 1.5$, $\La_\pi=1.5$\\
$g_{\pi NN^*}^2/4\pi = 0.9$ & $\La_{N^*} = 1.9$, $m_{N^*}=1.54$ \\
$f_{\pi N\De}^2/4\pi = 0.33$, $z=-0.2$ & $\La_\De = 1.4$, $m_\De=1.24$\\
$g_{\si NN}g_{\si\pi\pi}/4\pi = 0.1$ & $\La_\si = 1.1$, $m_\si=0.55$\\
$g_{\rho NN}g_{\rho\pi\pi}/4\pi = 2.5, \kappa_\rho=3.7
$& $\La_\rho = 1.1$, $m_\rho=0.77$
\end{tabular}
\vskip4mm
\caption{The model parameters which were ajusted to reproduce
the $\pi N$ phase-shifts.
Only the physical (renormalized) values of parameters are given.}
\end{table}

We shall now corroborate the issue of the full relativistic covariance  
in an explicit calculation of the self-energy contribution in our model
versus the models based on other relativistic QP equations for $\pi N$ 
system \cite{PeJ91,GrS93}.

The field theoretical nucleon self-energy in the models based on the BS
equation can be written,
after the partial-wave decomposition, as follows,
\beq
\eqlab{self1}
\Si_{r}(P_0) = \frac{i}{\pi}
\int\limits_{-\infty}^\infty\!\frac{\dd k_0}{2\pi}
\int\limits_0^\infty\! \frac{\dd\rk}{2\pi}\,\rk^2\,
\sum_\rho G^{(\rho)}(\rk,k_0;P_0)\,\Phi_r^{(\rho)}(\rk,k_0;P_0),
\eeq
where $G^{(\rho)}$ is the $\pi N$ propagator
and the interaction function $\Phi_r^{(\rho)}$ represents the
$\pi NN$ vertex contributions with $\rho$ characterizing the
$\rho$-spin\cite{tlh} of the intermediate state and $r$ characterizing
parity.
Furthermore, the integration variable $k_0$ is the relative-energy
variable, defined as
\beq
k_0 = ( p_{\N 0}\,\hat{\w} -  p_{\pi 0}\,\hat{E})/P_0,
\eeq
where $p_{\N 0}$ ($p_{\pi 0}$) is the 0-th component of the
nucleon (pion) 4-momentum, and
$
\hat{\w}=(P_0^2 - m^2 + m_\pi^2)/2P_0,\,
\hat{E}=(P_0^2 + m^2 - m_\pi^2)/2P_0.
$
Then  $G^{(\rho)}$ has the following explicit form,
\beq
G^{(\rho)}(k,k_0;P_0) =  \rho \left\{(\hat{E} + k_0 - \rho E_k+i\eps)
[ (\hat{\w}-k_0)^2 - \w_k^2-i\eps ]\right\}^{-1},
\eeq
which clearly indicates the position of its poles in  $k_0$ plane:
\beq
\eqlab{down}
k_0^{(\pi \pm)} = \hat{\w} \mp (\w_k + i\eps), \,\,\,\,
k_0^{(\N \pm)} = -\hat{E}  \pm ( E_k - i\eps),
\eeq
with $E_k = \sqrt{m^2+\rk^2},$ $\w_k = \sqrt{m_\pi^2+\rk^2}$.
Also, $\Phi_r^{(\rho)}$ has poles
associated with those of the potential and the form factors\footnote{In 
the lowest order
the r.h.s of \Eqref{self1} is simply the one $\pi N$ loop
with $\Phi_r^{(\rho)}$ the product of the two bare $\pi NN$ vertices.}.

In the ET approximation we have, instead of \Eqref{self1},
\beq
\eqlab{self2}
\Si_{r}(P_0) \stackrel{ET}{=} \frac{1}{\pi}
\int\limits_0^\infty\! \frac{\dd\rk}{2\pi}\,\rk^2\,
\sum_\rho G_{\rm ET}^{(\rho)}(\rk;P_0)\,\Phi_r^{(\rho)}(\rk,0;P_0),
\eeq
where
\beq
G_{\rm ET}^{(\rho)}(\rk;P_0)=
i\!\int\limits_{-\infty}^\infty\!\frac{\dd k_0}{2\pi}\,
G^{(\rho)}(\rk,k_0;P_0)=   -\rho \left\{2\w_k 
(-\rho P_0 + E_k + \w_k + i\eps)\right\}^{-1}.
\eeq

In \Figref{selff} we plot the results of the calculation
of the nucleon self-energy to one $\pi N$ loop, using 
the pseudo-vector coupling and the form factors described above.
As \Figref{selff} shows, within the ET approximation the 
results for $\Si_{+} (P_0)$ and $\Si_{-} (-P_0)$ are exactly the same.
As an example of a QP prescription which
violates the full covariance we may consider the spectator (or Gross) 
equation, where one of the particles in the loop is on its mass shell.
Formally, in doing the $k_0$ integral the contribution of only one
pole is taken into account: it can be either $k_0^{(\pi+)}$ in \Eqref{down}
 --- the pion spectator,
or $k_0^{(N+)}$   --- the nucleon spectator. The ``pion spectator''
was preferred by Gross and Surya\cite{GrS93}, and we have also used this choice
to carry out the calculation presented in \Figref{selff}.
The figure clearly shows that
the desirable identity \Eqref{posneg} is not satisfied for the
spectator model. Of course, the size of the violation is model-dependent,
but is unlikely to vanish for all energies in any model. Similar
violations arise in the equations used by Pearce and Jennings\cite{PeJ91}.
In particular, their choice of the Blankenbecler-Sugar propagator has this problem due to the presence of the positive-energy projection operator,
which, again, leaves out the contribution of the negative-energy pole.
In contrast,
the ET approximation satisfies the identity
\Eqref{posneg} exactly, thus preserving the Lorentz structure given in
\Eqref{form}.

\begin{figure}[p]
\epsffile{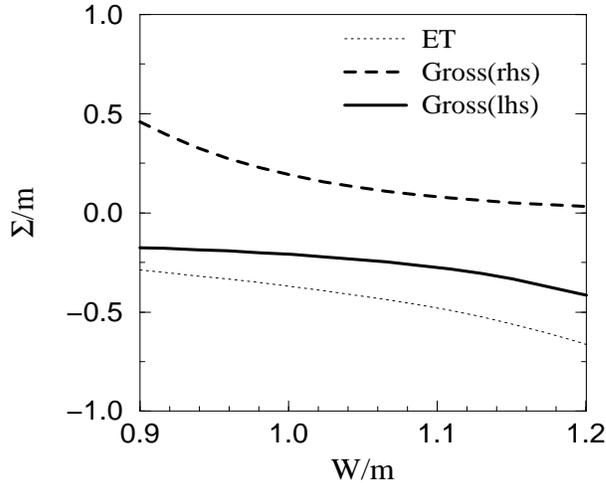}
\caption[F2]{ The r.h.s.\ and l.h.s.\ of \Eqref{posneg}
for the  one $\pi N$-loop using
the Gross prescription (pion spectator model). For the ET approximation
they are found to be identical,  given  by the dotted line.
Both the self-energy and the energy, $W=P_0$ are in
units of the nucleon mass.}
\figlab{selff}
\end{figure}

Note that all of the considered models \cite{PeJ91,GrS93,PaT97}
maintain the exact two-particle
(elastic) unitarity. They all thus give the following expression
for the imaginary part of the nucleon self-energy,
\beq
{\rm Im} \Si_{r}(P_0) = - \frac{q\hat{E}}{4\pi P_0}\,
\Phi_r^{(+)} (q, 0;P_0)\,\th(q^2),
\eeq
where $q$ the center-of-mass momentum, $\th$ is the step function.
In employing the ET prescription,
we could have for example tried to remove
one of the poles of the $\pi N$ propagator. 
However, in doing so we loose the two-particle
unitarity of the equation. Thus the condition
of elastic unitarity and full Lorentz covariance
clearly restricts  the form of allowed ET approximation.

Since s-channel singularities are present in our force model, we apply a renormalization procedure. Considering for instance the
dressed nucleon propagator \eref{dprop}, we see that the positive and
negative bare mass singularities in \Eqref{propcms} are shifted equally
in view of \Eqref{posneg}, yielding poles in the dressed propagator at
$P_0=\pm m$, $m$ being the physical nucleon mass.
Consequently, a standard renormalization constant $Z_2$ which transforms like
a scalar can be used. If \Eqref{posneg} does not hold, such a
renormalization procedure breaks down. In particular, it
would require non-covariant subtraction constants.

Hence, we may conclude that 
the violation of identity \eref{posneg}
leads to the difference in renormalization of the positive and negative
energy states. This obviously signals the violation of CPT symmetry. 
The non-covariance found in some QP prescriptions 
is indeed related to the violation of certain discrete Lorentz symmetries.
One can show that invariance under charge conjugation,
parity and time reversal implies the following transformation properties
for the spin-1/2 particle self-energy,
\bea
&&C: C \Si(p_0,\vec{p}) C^{-1} = \Si^T (-p_0,-\vec{p}),
\nonumber\\
&&P:  \ga_0 \Si(p_0,\vec{p}) \ga_0 = \Si (p_0,-\vec{p}),
\nonumber\\
&&T : \ga_0 C^{-1} \ga_5 \Si(p_0,\vec{p})\ga_5 C \ga_0 =
\Si^T (p_0,-\vec{p}),
\nonumber\\
&&CPT:  \ga_5 \Si(p_0,\vec{p}) \ga_5 = \Si(-p_0,-\vec{p}),
\nonumber
\eea
where superscript $T$ means transposition in the $\ga$-matrix space
and $C$ is the charge conjugation matrix
(note that here CPT and C imply essentially the same condition, since
$\ga_5$ and the $C$-matrix act very similar on $\ga_\mu$: $\ga_5 \ga_\mu \ga_5
 = -\ga_\mu,$ $C \ga_\mu C^{-1} = -\ga_\mu^T$).
From this we see that the violation of \Eqref{posneg} appears
as the violation of the charge conjugation and, as a consequence,
CPT symmetry.

In conclusion, we have presented a relativistic unitary model
for $\pi N$ interaction based on the ET type of formulation  
which is consistent also with the discrete Lorentz symmetries, in contrast
to the models based on other relativistic 3-D formulations. The violations
of the discrete Lorentz symmetries, such as charge conjugation and CPT,
 occur in the most of the 3-D formulations because 
of the difference in the treatment of positive and negative energy states. 
These violations are reflected in the violation
of the Lorentz structure, \ie dependences on not only the standard
covariants. Therefore, in these cases, one should be careful to use
covariant arguments to construct the transformation properties of 
the calculated amplitudes, which would be needed when we want to incorporate
the basic interaction in more particle systems. 
One should also be careful in exploiting the usual Ward identities, low-energy theorems, power-counting arguments.


\end{document}